\documentclass[aps,prl,reprint]{revtex4-2}

\usepackage{graphicx}
\usepackage{epstopdf, epsfig}
\usepackage{amsmath}

\usepackage{hyperref}

\usepackage[normalem]{ulem}
\usepackage{color}

\usepackage{dcolumn}% Align table columns on decimal point
\usepackage{bm}% bold math
\usepackage[mathlines]{lineno}% Enable numbering of text and display math
\usepackage{tikz}
\definecolor{magenta_1}{rgb}{0.64,0.08,0.18}
\definecolor{yellow_1}{rgb}{0.93,0.69,0.13}
\definecolor{lightOrange}{rgb}{0.9882, 0.8235, 0.6}
\definecolor{lightGray}{rgb}{0.898, 0.882, 0.902}

\newcommand{\orangeline}{\raisebox{2pt}{\tikz{\draw[-,orange,solid,line width = 0.9pt](0,0) -- (5mm,0);}}}

\newcommand{\blackline} {\raisebox{2pt}{\tikz{\draw[-,black,solid,line width = 0.9pt](0,0) -- (5mm,0);}}}

\newcommand{\lightorangeline}{\raisebox{2pt}{\tikz{\draw[-,lightOrange,solid,line width = 0.9pt](0,0) -- (5mm,0);}}}
\newcommand{\lightgrayline}{\raisebox{2pt}{\tikz{\draw[-,lightGray,solid,line width = 0.9pt](0,0) -- (5mm,0);}}}

\newcommand{\blackdash}{\raisebox{2pt}{\tikz{\draw[-,black,dashed,line width = 0.9pt](0,0) -- (5mm,0);}}}

\newcommand{\blackdotted}{\raisebox{2pt}{\tikz{\draw[-,black ,dotted,line width = 0.9pt](0,0) -- (5mm,0);}}}

\begin{document}

\preprint{APS/123-QED}

% \includepdf[pages=1-,pagecommand={\thispagestyle{plain}}]{<pdffile>}

\title{New insight into the spectra of turbulent boundary layers with pressure gradients}
\author{Ram\'on Pozuelo}
\author{Qiang Li}%
\author{Philipp Schlatter}
\author{Ricardo Vinuesa}
\affiliation{FLOW, Engineering Mechanics, KTH Royal Institute of Technology, SE-100 44 Stockholm, Sweden}

% \author{R. Pozuelo}
% \author{Q. Li}%
% \author{P. Schlatter}
% \author{R. Vinuesa}

% \author{N.~Arnaud}
% \author{J.~B'equilleux}
% \author{A.~D’Orazio}
% \affiliation{University of California at Irvine, Irvine, California 92697, USA}

\date{\today}

% Abstracts:  ≤ 600 characters
\begin{abstract}
With the availability of new high-Reynolds-number ($Re$) databases of turbulent boundary layers (TBLs) it has been possible to identify in detail certain regions of the boundary layer with 
more complex behavior. In this study we consider a unique database at moderately-high $Re$, with a near-constant adverse pressure gradient (APG) (Pozuelo {\it et al.}, {\it J. Fluid Mech.}, {\bf 939}, A34, 2022), and perform spectral analysis of the Reynolds stresses, focusing on the streamwise component. We assess different regions of the APG TBL, comparing this case with the zero-pressure-gradient (ZPG) TBL, and identify the relevant scaling parameters as well as the contribution of the scales of different sizes. 
The small scales in the near-wall region up to the near-wall spectral peak have been found to scale using viscous units.
In APG TBLs, the largest scales close to the wall have a better scaling with the boundary-layer thickness ($\delta_{99}$), and they are significantly affected by the APG.
In the overlap and wake regions of the boundary layer, the small energetic scales exhibit a good scaling with the displacement thickness ($\delta^*$) while the larger scales and the outer spectral peak are better scaled with the boundary-layer thickness. Also, note that the wall-normal location of the spectral outer peak scales with the displacement thickness rather than the boundary layer thickness. The various scalings exhibited by the spectra in APG TBLs are reported here for the first time, and shed light on the complex phenomena present in these flows of great scientific and technological importance.
\end{abstract}
% 248 words

\maketitle

% \section{Introduction}

Turbulent boundary layers (TBLs) play an essential role in a wide range of areas, including atmospheric flows, aircraft design or turbine blades. Assessing the effect of streamwise pressure gradients on TBLs is critical to completely understand these applications \cite{harun_monty2013, Maciel_2018}, and this is very challenging due to the complex effect of flow history on the local state of turbulence \cite{bobke2017, tanarro_2020, Kitsios2017}. Here we study adverse-pressure-gradient (APG) effects on TBLs on statistically two-dimensional flows subjected to a nearly-constant APG magnitude leading to near-equilibrium conditions \cite{Marusic_McKeon_2010, Nagib_2008}.
The turbulent character of the flow is analysed through the Reynolds decomposition \cite{Rey_decomp} of the fluid variables into a mean flow (averaged in time and the homogeneous spanwise direction) and a fluctuating component.
Through this statistical approach, the mean flow has been extensively studied and some universal laws have been established, such as the law of the wall \cite{vonKarman1931}, the defect law and log-law  \cite{millikan39}, together with various approaches to account for different flow configurations \cite{Luchini_2017}. 
One important tool for mathematical modeling of physical systems is scaling, which in the context of turbulent flows has been used for assessing the physics at higher Reynolds numbers~\cite{Hultmark_2012}, and also in combination with symmetry analyses~\cite{Oberlack_2022}.
The approach taken in this work is based on scaling different energetic regions of the boundary layer at high Reynolds numbers, and it can be useful for other multi-scale systems where there are different energy scales interacting with each other. In these cases, a marginal integration of the scales of various sizes can lead to understanding the relative relevance and sensitivity of the various scales.

In this study we assess the fluctuating components of APG TBLs, denoted by Reynolds stresses (RSs), where the streamwise normal component $\overline{u^2}$ is the most energetic. Note that $\overline{(\cdot)}$ denotes the average in time and the homogeneous spanwise direction.

In APG TBLs we deal mainly with two parameters: the Reynolds number and the APG magnitude, where the latter can also be a function of the streamwise position.
% \rev{It is interesting to find scaling factors for certain characteristics of the flow such that they can collapse with Reynolds number and/or APG magnitude, so as to further understand the physical behavior of the flow.}
Kitsios {\it et al.}~\cite{Kitsios2016, Kitsios2017} used an integral approach to self-similarity in order to obtain scaling factors for the Reynolds stresses, and compared the spectra of a ZPG and an APG. 
The integral scaling factor they used was the displacement thickness, which is defined as $\delta^*=\int_0^{\delta_{99}} (1 - U/U_{e}) \mathrm{d}y $, where $U$ is the mean streamwise velocity and $U_e$ is the velocity at the edge of the boundary layer ($y=\delta_{99}$). Note that $x$, $y$ and $z$ are the streamwise, wall-normal and spanwise coordinates, respectively. 
% \rev{The outer peak of all the RSs was shown to be at $y \approx 1.2\delta^*$ in Kitsios {\it et al.}~\cite{Kitsios2016} for their mild-APG simulation and at $y \approx \delta^*$ for their strong-APG case~\cite{Kitsios2017}. }
Following this RS scaling, Maciel {\it et al.}~\cite{Maciel_2018} and Sanmiguel Vila {\it et al.}~\cite{Sanmiguel_PRF} scaled the outer peaks of the RSs with $\delta^*$ and $\delta_{99}$ for several numerical and experimental databases with a large range of APG magnitudes. 
% \rev{ The former focused on non-equilibrium databases, such as the APG TBLs around wing profiles, and the latter showed results for experimental near-equilibrium TBLs developing on flat plates. }
In these studies, it was found that the wall-normal location of the outer peak of the RSs scales with $\delta^*$, making it a scaling independent of the Reynolds number and APG effects. 

Note that Maciel {\it et al.}~\cite{Maciel_2018} reflected on the $\delta^*$ scaling of the wall-normal location of the outer peak, $y_{\rm OP}$. Since $\delta^*/\delta_{99}$ decays slowly to zero when the Reynolds number tends to infinity, this would imply that the outer peak approaches the wall, therefore $\delta^*$ would only scale $y_{\rm OP}$, but not necessarily the size of the outer region.

Despite the different values for the wall-normal location of the outer peaks in the RSs for moderate and strong APGs documented in Ref.~\cite{Kitsios2017}, the spectral outer peak for the ZPG and the strong-APG cases were shown to be at the same wall-normal location: $y \approx \delta^*$. 
% \rev{
% Furthermore, the spanwise scale of the strong APG outer peak was reported to be at $\lambda_z \approx 2\delta^*$.
% }
In other studies where $\delta_{99}$ was used for scaling the outer region of the TBL \cite{Lee2017, bobke2017}, the outer-spectral-peak wavelength $\lambda_z$ appears to scale better in both ZPG and APG with $\delta_{99}$ than with $\delta^*$. 

% \rev{
% Most of the previous studies focused on short regions of interest and/or low Reynolds numbers, and the studies reaching higher Reynolds numbers were either non-equilibrium simulations or a result of a strong APG, resulting in a very pronounced development of the TBL. On the other hand, Pozuelo {\it et al.}~\cite{Pozuelo_JFM_22} conducted a well-resolved large-eddy simulation (LES) of a moderate-APG TBL which develops from ZPG conditions into a near-equilibrium APG with nearly-constant pressure-gradient magnitude. This unique database exhibits a long region of near-equilibrium flow at high Reynolds number, which enabled assessing the behavior of the near-wall and outer peaks of the streamwise Reynolds stress $\overline{u^2}$.
% }
Based on the work by Pozuelo {\it et al.}~\cite{Pozuelo_JFM_22}, in APG TBLs the wall-normal location and magnitude of the near-wall peak, denoted here by $y_{\rm IP}$ and $\overline{u^2}_{\rm IP}$ respectively, were found to increase with Reynolds number and APG magnitude using inner scaling. In this scaling, the viscous length $\ell_{\tau}=\nu/u_{\tau}$ and the friction velocity $u_{\tau}$ are employed, where $\nu$ is the fluid kinematic viscosity and $u_{\tau}=\sqrt{\tau_w/\rho}$ (where $\tau_w$ is the wall-shear stress and $\rho$ is the fluid density). The superscript `+'  will be used to denote inner scaling.
Regarding the outer peak, its wall-normal location $y_{\rm OP}$ was found to be approximately constant when scaled with either the displacement thickness $\delta^*$ or the boundary-layer thickness $\delta_{99}$, since for a moderate APG $\delta^*/\delta_{99}$ decreases slowly with the Reynolds number.
With the former length scale, $y_{\rm OP}/\delta^*$ appeared to be less influenced by the APG, obtaining an approximate value of 1.4, while using $\delta_{99}$, the locations varied with APG magnitude. 

% \rev{
% On the other hand, the ZPG case exhibits some energy in the outer region which did not develop into an outer peak of $\overline{u^2}$; however, the power-spectral density of this quantity exhibits an outer peak similar to that of the APG case.
% }
Analyzing the RS using the spanwise spectra with different scaling factors, we will assess similarities and differences with respect to the ZPG and the scaling properties of the RS. This has important implications in the context of the fundamental structure of turbulence.

% \section{Databases}
In this study we analyze two high-Reynolds-number databases of well-resolved LES of ZPG \cite{EAmorZPG} and APG TBLs \cite{Pozuelo_JFM_22}, focusing on the near-equilibrium region in the APG, {\it i.e.} for $Re_{\tau}>1000$. Note that $Re_{\tau}$ is the friction Reynolds number based on the friction velocity $u_{\tau}$, and can be written as $Re_{\tau}=\delta_{99}/\ell_{\tau}=\delta_{99}^+$.
The colours and line styles used for the two databases are shown in Table~ \ref{tab:PGcases}.
% ---------------------------------------------------
% ------------Databases and symbols------------------
% ---------------------------------------------------
\begin{table}
\centering
\begin{tabular}{lc||cr}
\textrm{PG } & \textrm{Colour} & \textrm{$Re_{\tau}$} & \textrm{Style} \\
\colrule
APG & $\orangeline$ or $\lightorangeline$ & 1000 & $\blackline$  \\
ZPG & $\blackline$ or $\lightgrayline$  & 1500 & $\blackdash$  \\
    &                & 2000 & $\blackdotted$ \\
\end{tabular}
\caption{(Left) Colors used to denote the contour lines of the two TBL databases and (right) line styles used to denote the analyzed profiles at different friction Reynolds numbers in the two TBL cases. }
\label{tab:PGcases}
\end{table} 

% ---------------------------------------------------
% \section{Spanwise 1D Spectra}
% \section{\label{sec:InnerPeak} Scaling:\protect\\ Inner Peak }
% ---------------------------------------------------

The spanwise power-spectral density of a Reynolds-stress component $\overline{u_iu_j}(y)$ is denoted as $\phi_{u_iu_j}(y,k_z)$, where $k_z$ is the spanwise wavenumber.
% \rev{
% and $y$ is the wall-normal coordinate.
% }
Note that the integral over all the scales is equal to the total RS. Using this property, we can use an integral to determine the marginal contribution of small or large scales to the total RS (marginal contribution of energy, MCE). This can be written as follows:
\begin{equation}
\mathrm{MCE} = \int_{k_{z,c}}^{\infty} \phi_{u_iu_j} \mathrm{d} k_z \ \bigg/ \int_{0}^{\infty} \phi_{u_iu_j} \mathrm{d} k_z,
\label{eq:cumsum}
\end{equation}
where $k_{z,c}$ is the cut-off wavenumber.
The spectra will be analysed through contours of the premultiplied power-spectral density $k_z\phi_{u_iu_j}$ and contours of the marginal contribution to the RS as a percentage. Instead of the wavenumbers $k_z$, the contours will be represented against the spanwise wavelength $\lambda_z=2\pi/k_z$.

The pre-multiplied spanwise spectra of the streamwise RS $k_z\phi_{uu}(y,\lambda_z)$ for both the APG and ZPG clearly show the near-wall and outer spectral peaks, as can be observed below in Fig.~\ref{fig:inner_kPSDz}(a). 
The location of these spectral peaks can be scaled with a length scale to collapse the wall-normal position, here denoted as $y_{s \rm IP}$ and $y_{s \rm OP}$, and another length scale can be used to scale the wavelength of the peaks $\lambda_{z,s \rm IP}$ and $\lambda_{z,s \rm OP}$.
In previous studies, the same scaling factor was used for the wall-normal location and for the wavelength $\lambda_z$; the novelty here is that we allow for a mixed scaling, {\it i.e.} the spanwise structures can have a different scaling than the wall-normal position where they are located, and Fig.~\ref{fig:inner_kPSDz}(b) shows the success of this approach.
A more complex analysis would be to determine the scalings for the energetic contours, since an energy scaling would be required for $k_z\phi_{uu}$, and the contours can be taken at a certain energy level or at a certain percentage of the peak or the maxima. For this last analysis the use of the MCE is useful. 

% \subsection{Peak values of the streamwise spectra}

\begin{figure}
\includegraphics[width=0.49 \textwidth]{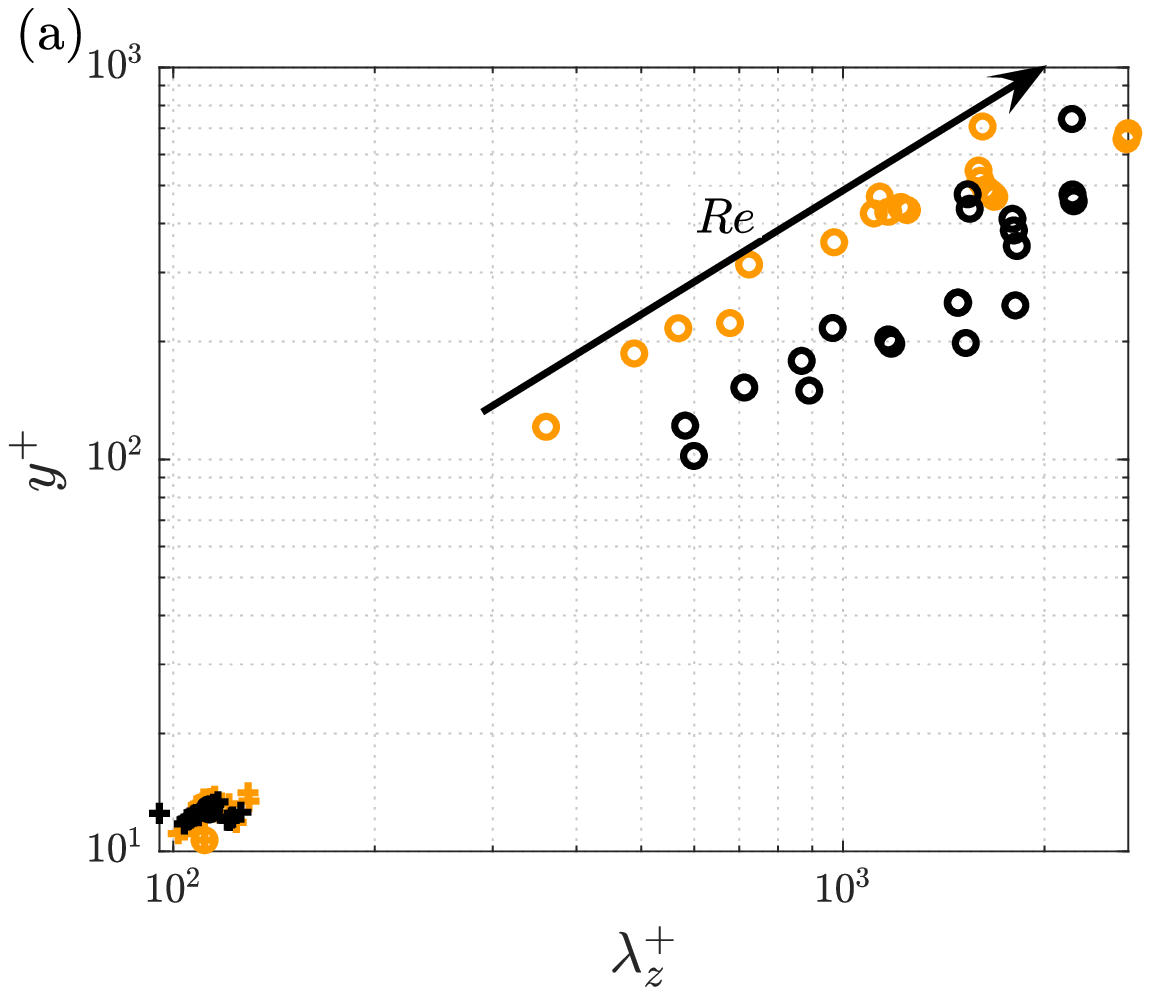}
\includegraphics[width=0.49 \textwidth]{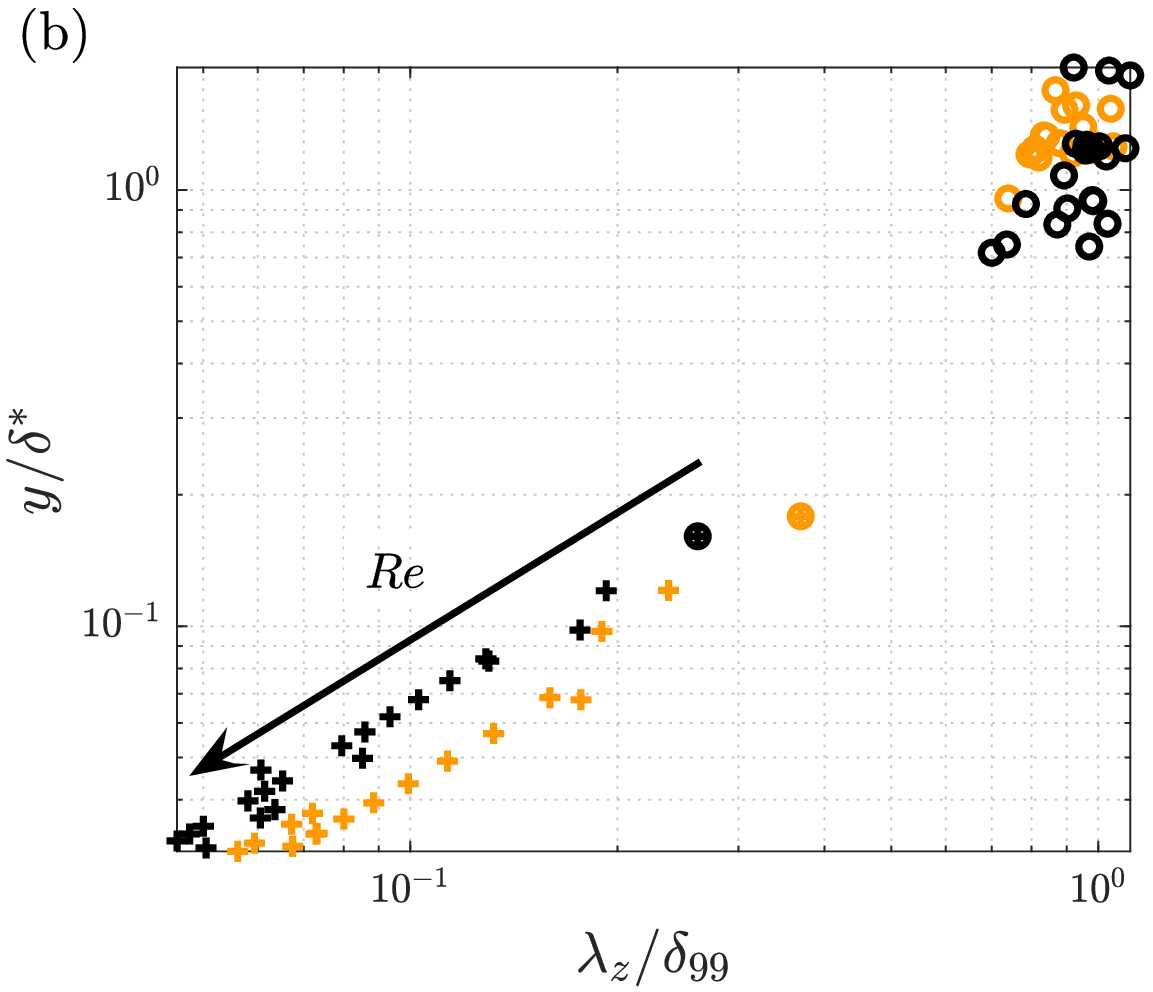}
\caption{ \label{fig:peaks} Representation of the peaks in the premultiplied streamwise power-spectral density $k_z\phi_{uu}$ in (a) inner scaling and (b) scaling with boundary-layer and displacement thicknesses. Crosses denote the inner peak (IP) and circles for the outer peak (OP). Note that in (a) the inner peaks collapse by scaling the wall-normal position $y$ and the wavelength $\lambda_z$ in viscous units, and in (b) the outer peaks collapse when scaling $y$ and $\lambda_z$ with $\delta^*$ and $\delta_{99}$, respectively. The black arrows show the direction of increasing Reynolds number.}
\end{figure}

The near-wall peak of $\overline{u^2}$ is closely connected with the spectral near-wall peak of $k_z\phi_{uu}$, thus it is natural to use the viscous scaling for this region as shown in Fig.~\ref{fig:peaks}(a) for a number of profiles between $Re_{\tau}=1000$ and 2000. This figure shows that the spectral inner peaks are located at approximately the same location for both APG and ZPG, at $y_{s \rm IP}^+\simeq 12.5$, without a clear trend due to the resolution taken for the spectra close to the wall.
The spanwise wavelenghts of the near-wall peak also collapse using inner scaling to a value $\lambda_{z,s \rm IP}^+ \approx 112$.

The outer peak of the RS is related with the spectral outer peak, and Fig.~\ref{fig:peaks}(b) shows that this spectral peak collapses in the premultiplied spectra. In particular, the wall-normal location of the outer peak scales with $\delta^*$ as in Ref.~\cite{Kitsios2017}, while the spanwise wavelength $\lambda_z$ scales with $\delta_{99}$ \cite{Lee2017, bobke2017}. Again, due to the resolution of the spectral data, the values are slightly scattered. 
Our results show that the average $y_{s \rm OP}/\delta^*$ is 1.3 for the APG and 1.1 for the ZPG. The average spanwise wavelenghts $\lambda_{z, s \rm OP}/\delta_{99}$ are 0.91 for the APG and 0.96 for the ZPG.
Considering the collapse of the inner and outer peaks in their respective scalings, then $y^+_{s \rm IP}/\lambda^+_{z, s \rm IP}=C_1$ and ($y_{s \rm OP}/\delta^*)/(\lambda_{z, s \rm OP}/\delta_{99}) = C_2$ are constant values. With the values above we obtain $C_1 \approx 0.1$ (for both ZPG and APG), and $C_2 \approx$ 1.43 for APG and 1.15 for ZPG.
Then the curves that the spectral outer peaks follow in Fig.~\ref{fig:peaks}(a) follow Eq.~(\ref{eq:curve_sOP}), and the spectral inner peaks in Fig.~\ref{fig:peaks}(b) follow Eq.~(\ref{eq:curve_sIP}). Note that $f(Re, \rm PG) = \delta^*/\delta_{99}$, where ${\rm PG}$ denotes the pressure-gradient magnitude. 
\begin{equation}
    y_{s \rm OP}^+ = \frac{y_{s \rm OP}/\delta^*}{\lambda_{z, s \rm OP}/\delta_{99}} \frac{\delta^*}{\delta_{99}} \lambda_{z, s \rm OP}^+  = C_2 f(Re, \rm PG) \lambda_{z, s \rm OP}^+
    \label{eq:curve_sOP}
\end{equation}
\begin{equation}
        \frac{y_{s \rm IP}}{\delta^*} = \frac{y_{s \rm IP}^+}{\lambda^+_{z, s \rm IP}} \frac{\delta_{99}}{\delta^*} \frac{\lambda_{z, s \rm IP}}{\delta_{99}}  = C_1 \frac{1}{f(Re, \rm PG)} \frac{\lambda_{z, s \rm IP}}{\delta_{99}}.
        \label{eq:curve_sIP}
\end{equation}

Since $\delta^*$ and $\delta_{99}$ are outer scales that can be measured in experiments~\cite{diagnostic_Vinuesa}, Eqs.~(\ref{eq:curve_sOP}) and (\ref{eq:curve_sIP}) can be useful to estimate the location of the inner and outer spectral peaks in experiments where the spectra cannot be easily measured. Also note that our results for $y_{\rm OP}$ reported in Pozuelo {\it et al.}~\cite{Pozuelo_JFM_22} are in agreement with Refs.~\cite{Kitsios2017, Maciel_2018, Sanmiguel_PRF}, while $y_{s \rm OP}$ also scales with $\delta^*$ as in Kitsios {\it et al.}~\cite{Kitsios2017}, in a case where the APG is at the verge of separation. The difficulties to distinguish trends using $\delta^*$ or $\delta_{99}$ arise from the fact that we have a mild APG, where $\delta^*/\delta_{99}$ decreases very slowly with Reynolds number.
For larger APG magnitudes, the decay of $\delta^*/\delta_{99}$ is larger with Reynolds number, as can be observed when comparing the results for an APG at the verge of separation shown in Ref.~\cite{Kitsios2017} and the mild-APG in this work. The values for $y_{s \rm OP}/\delta^*$ are similar in both studies, confirming the $\delta^*$ scaling.
As for the wavelength $\lambda_{z,s \rm OP}$, the scaling with $\delta_{99}$ leads to similar trends for the ZPG and the APG simulations as those documented in Refs.~\cite{Lee2017, bobke2017}.

The magnitudes of the peaks are also relevant, since they indicate the energy scale that can be used to scale the energy contours of different regions.
Pozuelo~{\it et~al.}~\cite{Pozuelo_JFM_22} reported that $\overline{u^2}^+_{\rm IP}$ grows with Reynolds number and APG, however, we have observed that the values of the spectral near-wall peak appear to be constant with the Reynolds number and the value for the ZPG case is slightly higher ($\overline{u^2}^+_{s \rm IP}=3.6$) than that of the APG case ($\overline{u^2}^+_{s \rm IP}=3.4$).

Although the ZPG data does not exhibit an outer peak in the RS, the outer region scales with the edge scaling as well as the APG \cite{Pozuelo_JFM_22}; here we assess whether the spectral outer-peak magnitude scales using the square of the edge velocity $U_{e}^2$, 
which is determined in APG TBLs following the methodology described in Ref.~\cite{diagnostic_Vinuesa}.
Our results show that the variation of the values obtained in the outer peak of $k_z\phi_{uu}/U_e^2$ was too large to determine any trend, however, the average level is higher for the APG ($\simeq 4.5 \times 10^{-3}$) than for the ZPG case ($\simeq 3.4 \times 10^{-3}$).

% \subsection{Near-wall region}
% \rev{
% Since the inner peaks of the RS and the spectra scale properly in inner scaling, 
% }
In Fig.~\ref{fig:inner_kPSDz}(a) we focus on the region around the spectral inner peak scaled in viscous units.
It can be observed that the contours with $\lambda_z^+ < 400$ which are below $y^+ = 40$ scale properly in viscous units. The magnitude of the near-wall peak is lower for the APG, however, the energy around it is similar for the APG and the ZPG cases.
The MCE blue contours representing $20\%$ of the total $\overline{u^2}$ show that the smallest scales provide a similar contribution close to the wall at different Reynolds numbers and for both APG and ZPG. The $50$ and $80\%$ contours show the more prominent role of the large scales in this region with the APG, and also at higher Reynolds numbers.
The green lines ($\rm MCE=50\%$) below $y^+=10$ at $\lambda_z^+ \approx 140$ for the ZPG and $\lambda_z^+ \approx 170$ for the APG, are located in a region where the APG and ZPG have similar energy levels. The spectral inner peak is located between the blue and green lines, implying that these scales around the inner-peak add up to $30\%$ of the energy. The light green lines are at a larger $\lambda_z^+$, showing that the APG spectral inner peak is slightly less energetic than in the ZPG case. 
The fact that $\overline{u^2}^+_{\rm IP}$ is larger at higher Reynolds numbers is explained by the impact of the wider scales (larger $\lambda_z$), which exhibit good scaling with $\delta_{99}$ as it can be seen in Fig.~\ref{fig:inner_kPSDz}(b). The red lines indicate that the large scales contributing up to $20\%$ of the total RS collapse for different Reynolds numbers all across the TBL and the APG largest scales are more energetic than those for the ZPG case.

\begin{figure}
\includegraphics[width=0.49 \textwidth]{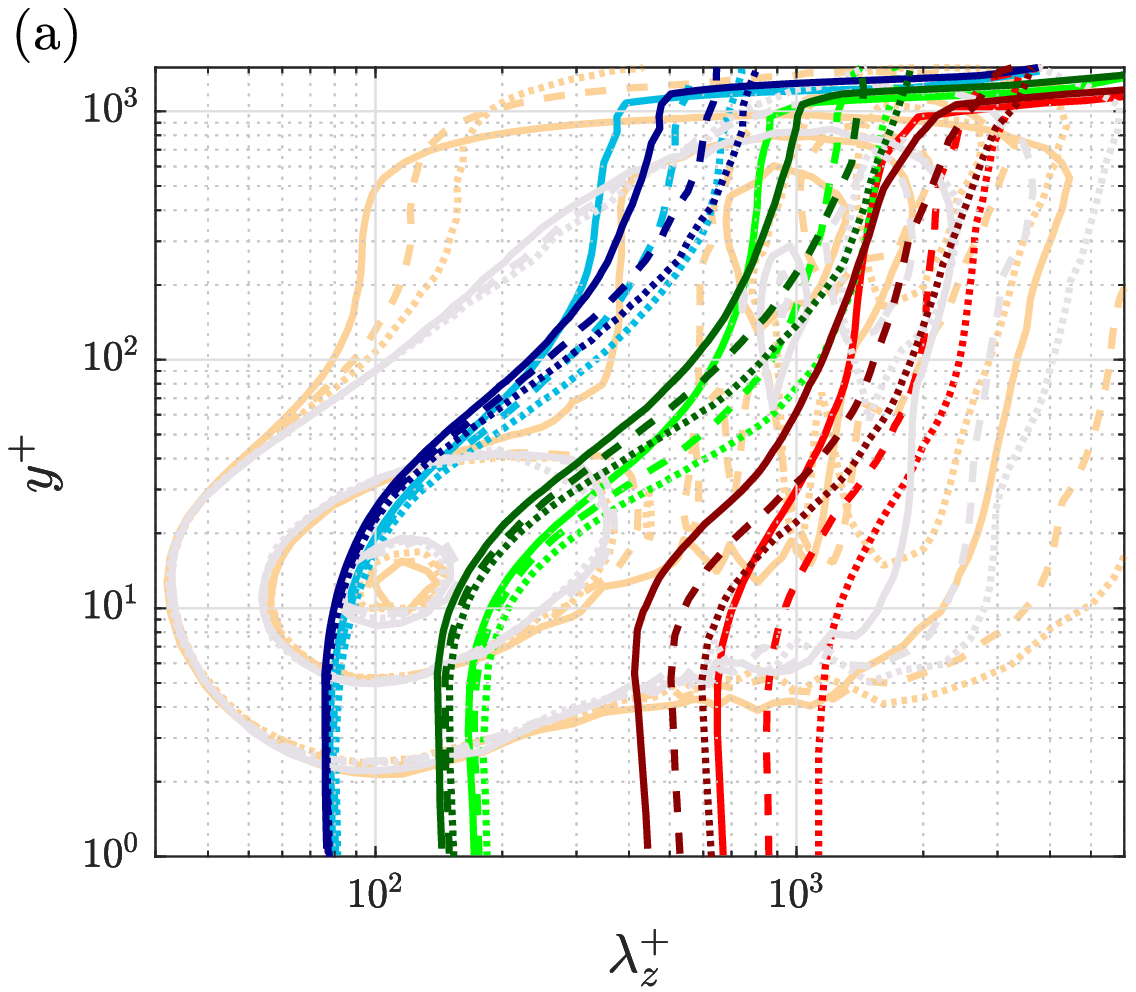}
\includegraphics[width=0.49 \textwidth]{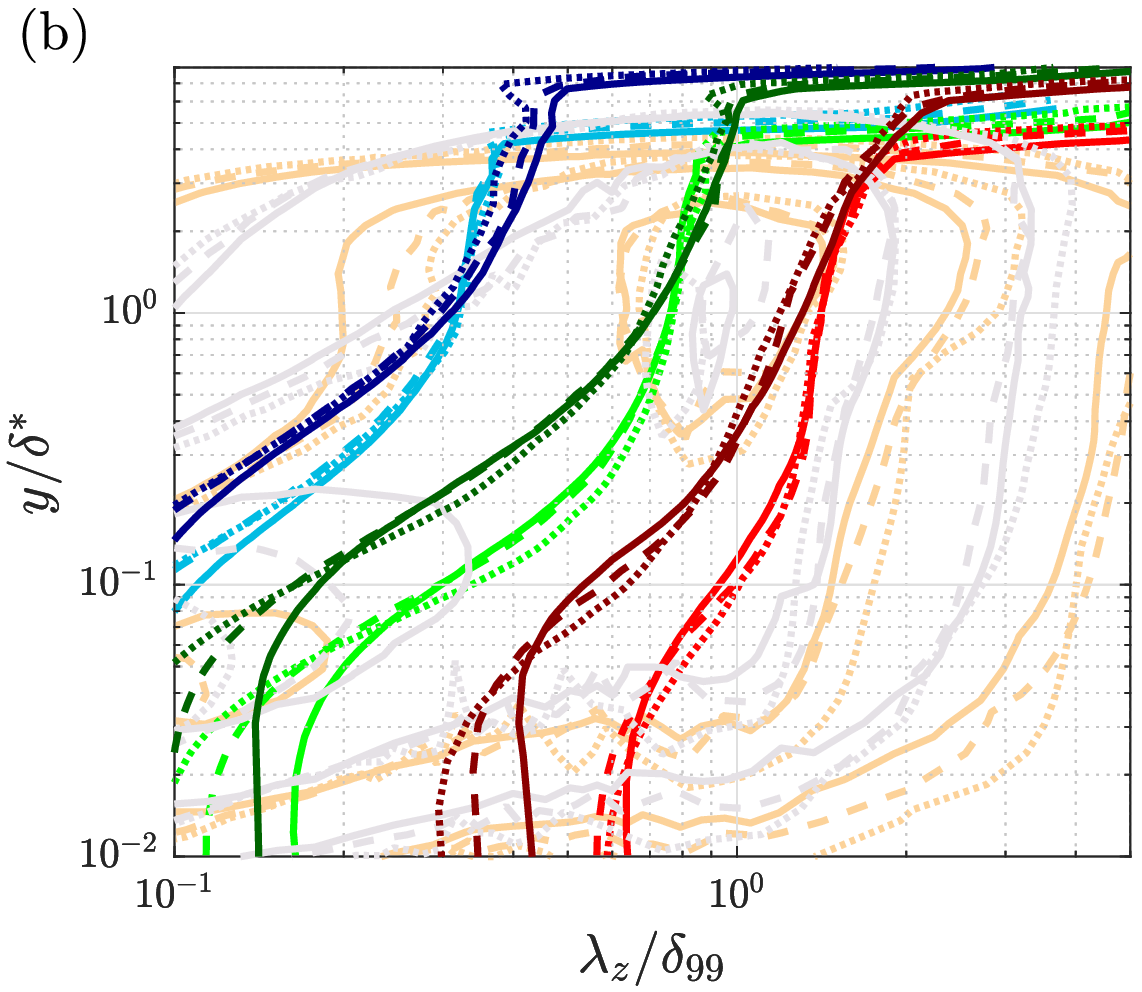}
\caption{ \label{fig:inner_kPSDz} Contours of the premultiplied spanwise power-spectral density of the streamwise RS, (a) scaled with the friction velocity, $k_z\phi_{uu}/u_{\tau}^2$, and (b) scaled with the edge velocity, $k_z\phi_{uu}/U_{e}^2$, both shown in lighter colors according to Table~\ref{tab:PGcases}. Contours taken at energy levels 0.5, 1.8 and 3.2 in panel (a). The levels for contours in panel (b) are: $[0.4, 1.2, 3] \times 10^{-3}$. The blue, green and red lines represent the size of the scales where the accumulated energy from the smallest scales add up to 20, 50 and $80\%$ of the total RS at that wall-normal location, respectively; for these lines, the light colors are used for the APG, while the dark colors are used for the ZPG.}
\end{figure}

% \subsection{Outer region}

In the outer region of the APG TBL it is also possible to differentiate two regions, the first one containing small energetic scales and another region of large scales surrounding the spectral outer peak.
As observed in Fig.~\ref{fig:peaks}(b), using $y/\delta^*$ and $\lambda_z/\delta_{99}$, the premultiplied spectral outer peak collapses for both APG and ZPG at different Reynolds numbers, although the value of the spectral outer peak is higher in the APG than for the ZPG case.
Since the outer scaling with $U_e^2$ yields a good collapse of $\overline{u^2}$ in the outer region for both cases \cite{Pozuelo_JFM_22}, the premultiplied spectra will be scaled with $U_e^2$, and the contours will be taken at a similar energy level.

%----------------------------------------------------------------------------------
%----------------------------------------------------------------------------------

In Fig.~\ref{fig:inner_kPSDz}(b), the outer spectral peak at different Reynolds numbers are located around the same location for both APG and ZPG and its scales are of a similar size, however, the APG energizes the scales around that peak.
% \rev{
% Wide scales of size $\lambda_z=4 \delta_{99}$ extend their influence across the TBL in the APG, while for the same energetic level the scales in the ZPG are of size $\lambda_z=3 \delta_{99}$.
% }
The large scales above $y=\delta^*$ contribute in a similar proportion to the RS (up to their respective $\delta_{99}/\delta^*$ values), independently of the Reynolds number or the APG (see red and green lines), while some differences start to be observed for the smaller scales (blue lines), since the APG contains more energy in smaller scales in this region. 
Between $y=0.1\delta^*$ and $y=\delta^*$, the large scales contribute in a similar way at different Reynolds numbers, but the relevance of the large scales is greater in the APG than in the ZPG. 
\begin{figure}
\includegraphics[width=1 \columnwidth]{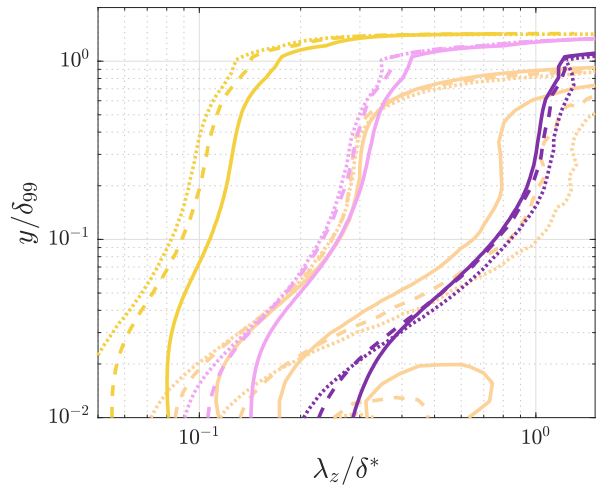}
\caption{ \label{fig:cont_cumsum_d99_dstar} Outer-scaled premultiplied power-spectral density of the streamwise RS, $k_z\phi_{uu}/U_e^2$. Contours taken at $[0.4, 1.2, 3] \times 10^{-3}$. The yellow, pink and purple lines are the MCE contours at 0.1, 2.2 and $15\%$, respectively.}
\end{figure}

In Fig.~\ref{fig:cont_cumsum_d99_dstar} we focus on the energetic small scales in the outer region of the APG. The energetic contours of $k_z\phi_{uu}/U_e^2$ are taken at the same levels as those in Fig.~\ref{fig:inner_kPSDz}(b). The smaller energy level (light orange lines) collapses for scales of size $\lambda_z \approx 0.3 \delta^*$ where the MCE is $2.2\%$ (pink lines). 
Less energetic contours were investigated to assess whether the scaling $\lambda_z/\delta^*$ is still valid for smaller scales. This resulted in a behaviour similar to what the yellow MCE contours show. Other scalings such as $\delta_{99}$ lead to further distancing of the energy contours for scales $\lambda_z < 0.3\delta^*$.

This is the first time that the energy of the small scales in the outer region of an APG TBL is assessed in detail in terms of their scaling and their MCE.
From the MCE contours it can be observed that the smallest scales in the outer region add up to $2.2\%$ below $\lambda_z = 0.3 \delta^*$ and $15\%$ for scales $\lambda_z < \delta^*$.
Although the energy contour levels around $\lambda_z = \delta^*$ do not exhibit any collapse with the Reynolds number, the MCE purple lines do, implying that the scales between $\lambda_z = 0.3 \delta^*$ and $\lambda_z = \delta^*$ (pink and purple lines) contribute in a similar way to the total RS.
The lack of collapse in the energy contours of the premultiplied spectra would in principle indicate that there is no similarity between the scales in this region; however, the MCE reveals that their marginal contribution is in fact very similar.

This small-scale energy in the outer region is not present in the ZPG, and the MCE contours do not exhibit any collapse, thus we did not include those contours in Fig.~\ref{fig:cont_cumsum_d99_dstar}.
%----------------------------------------------------------------------------------
%----------------------------------------------------------------------------------
The Reynolds shear stress was also analysed. Its spectra in the near-wall region also scales in viscous units and even exhibits an inner peak in both APG and ZPG located at $y^+ \simeq 25$ and $\lambda_z^+ \simeq 105$. The energy level is similar in APG and ZPG using $u_{\tau}^2$. 
The Reynolds shear-stress spectrum also develops an outer spectral peak in both APG and ZPG, which is centered around $y/\delta^*=1.4$ and $\lambda_z/\delta_{99} = 0.9$. The energy of the outer spectral peak in the APG is almost twice the energy contained in its inner peak, while in the ZPG the values of the energy in its inner and outer peaks is similar.
% ---------------------------------------------------
% \section{Conclusions}
% \section{\label{sec:InnerPeak} Scaling:\protect\\ Inner Peak }
% ---------------------------------------------------

The analysis of the $\overline{u^2}$ spectrum for APG TBLs conducted here included using different length scales for the wall-normal location and the spanwise wavelength.
The inner region in TBLs with moderate APG exhibits the classical results reported for other wall-bounded flows, where the viscous scaling leads to marginal variations with Reynolds number.
The wall-normal location of the outer spectral peak has been shown to scale with $\delta^*$ and its wavelength scales with a different outer scale: $\delta_{99}$. 
While the MCE contours are also useful to observe the limits of the inner and outer scalings, they also reveal that the small-scale energy present in the outer region of APG TBLs scales with $\delta^*$. The scales between $\lambda_z=0.3\delta^*$ and $\lambda_z=\delta^*$ may have larger energy levels at higher Reynolds numbers, but their MCE shows that they contribute in a similar proportion to the total outer energy level.

%Acknowledgements
RV acknowledges the financial support provided by the Swedish Research Council (VR).
PS and QL were supported by the KAW Foundatation.
The computations and data handling were enabled by resources provided by the Swedish National Infrastructure for Computing (SNIC), partially funded by the Swedish Research Council.

% \end{document}
\bibliography{apssamp}% Produces the bibliography via BibTeX.

\end{document}